# $Cr_3X_4$ (X=Se, Te) monolayers as new platform to realize robust spin filter, spin diode and spin valve

Qihong Wu,[1] Rongkun Liu,[1] Zhanjun Qiu,[1] Dengfeng Li,[1] Jie Li,[2] Xiaotian Wang,[3,*] and Guangqian Ding[1,*]

[1]*School of science, Chongqing University of Posts and Telecommunications, Chongqing 400065, China.*
[2]*Key Laboratory for the Physics and Chemistry of Nanodevices, School of Electronics, Peking University, Beijing 100871, China.*
[3]*School of Physical Science and Technology, Southwest University, Chongqing, 400715, China.*
E-mails: xiaotianwang@swu.edu.cn; dinggq@cqupt.edu.cn;

Two-dimensional ferromagnetic (FM) half-metals are promising candidates for advanced spintronic devices with small-size and high-capacity. Motivated by recent report on controlling synthesis of FM $Cr_3Te_4$ nanosheet, herein, to explore the potential application in spintronics, we designed spintronic devices based on $Cr_3X_4$ (X=Se, Te) monolayers and investigated their spin transport properties. We found that $Cr_3Te_4$ monolayer based device shows spin filtering and dual spin diode effect when applying bias voltage, while $Cr_3Se_4$ monolayer is an excellent platform to realize a spin valve. The different transport properties are primarily ascribed to the semiconducting spin channel, which is close to and away from the Fermi level in $Cr_3Te_4$ and $Cr_3Se_4$ monolayers, respectively. Interestingly, the current in monolayer $Cr_3Se_4$ based device also displays a negative differential resistance effect (NDRE) and a high magnetoresistance ratio (up to $2\times10^3$). Moreover, we found thermally induced spin filtering effect and NDRE in $Cr_3Se_4$ junction when applying temperature gradient instead of bias voltage. These theoretical findings highlight the potential of $Cr_3X_4$ (X=Se, Te) monolayers in spintronic applications and put forward realistic materials to realize nanosale spintronic device.

## I. INTRODUCTION

Spintronic devices, utilizing the spin of electrons to transport information, provide great opportunities for next-generation quantum information technology. An important approach to realize spin-based devices is the high-efficient spin injection from a magnet to a semiconductor. In this regard, ferromagnetic (FM) half-metals, which has a metallic state in one spin channel while a semiconducting or insulating state in other spin channel, provide completely spin polarized electrons around the Fermi level and is thereby an ideal spin injection source [1-3]. Indeed, the discovery of FM half-metals opens the way for numerous application including spin filters, spin diodes, spin valves, spin field effect transistors as well as spin Seebeck devices [4-8]. In recent years, two-dimensional (2D) FM materials as promising candidates for nanoscale spin-based devices have became a central topic of spintronic research. In 2015, Wu *et al.* reported from first-principles calculations that monolayer $MnN_2$, $YN_2$ and $NbN_2$ are FM metal, FM half-metal and FM semiconductor, respectively [9]. Feng *et al.* revealed the FM order in $FeCl_2$, $CoCl_2$ and

NiCl$_2$ monolayers with both 1T and 1H phase, and most of them are FM half-metals with large half-metallic gap [10]. In 2019, Zhang *et al*. predicted that 2D Cr$_3$X$_4$ (X=Se, Te) are FM half-metals with high Curie temperature [11]. Moreover, the FM half-metallicity of 2D Mn$_3$X$_4$ (X=Se, Te) was predicted by Chen *et al* [12]. In general, the successful prediction of 2D FM materials with high Curie temperature and intrinsic half-metallicity enriches the material candidates for spintronic device.

Designing spintronic devices based on realistic FM materials is an important approach to possible spin-based applications. The development of quantum transport approach opens the door for simulating transport properties of artificial devices. For example, Han *et al*. designed a GaAs/CoFeMnSi heterosutructure and a CoFeMnSi/GaAs/CoFeMnSi magnetic tunnel junction (MTJ) and found perfect spin diode effect and large tunnel magnetoresistance (TMR) [13]. Based on 2D YN$_2$ FM monolayer, Li *et al*. found dual spin diode effect and spin Seebeck effect in the constructed device [14]. In 2018, Wu *et al*. proposed a boron-nitrogen nanotube as a superior platform to realize perfect spin Seebeck effect and thermal spin filter [15]. To realize a large magnetoresistance in a spin valve, FM half-metals, with a large half-metallic gap and with the Fermi level lies in the centre of gap, are superior candidates for this target. Taking 1T-FeCl$_2$ monolayer as an example, perfect spin filtering effect, NDRE, and large magnetoresistance ratio up to $1.34 \times 10^5$ was discovered in the device as proposed by Feng *et al* [10]. Afterwards, they also found high TMR ($6.3 \times 10^3$) in a designed 1T-FeCl$_2$/2H-MoS$_2$/1T-FeCl$_2$ van der Waals junction [16]. In addition to theoretical design, experimental reported MTJ device, a van der waals junction formed by four layer CrI$_3$ sandwiched by monolayer graphene, has a high TMR approaching to $5.7 \times 10^3$ [17].

Although numerous 2D FM materials and their spin-based devices have been theoretically identified and realized, experimental evidence for these materials and devices has been seldom reported. In recent years, experimental synthesized 2D FM materials, such as Cr$_2$Ge$_2$Te$_6$ monolayer [18], CrI$_3$ monolayer [19], Fe$_3$GeTe$_2$ monolayer [20] and 1T-FeCl$_2$ monolayer [21], spark the development of nanoscale spintronic devices. However, these intrinsic FM monolayers have low Curie temperature and can not meet most spintronic applications. In order to achieve nanocale spin-based devices working under room temperature, intrinsic 2D half-metals with high Curie temperature is important candidates that should be highly sought out. Cr$_3$X$_4$ (X=Se, Te) monolayers are intrinsic FM half-metals with high Curie temperature (up to 370K for Cr$_3$Se$_4$ and 460 K for Cr$_3$Te$_4$), as theoretically predicted by Zhang *et al* [11]. Very recently, it is interesting that air-stable FM Cr$_3$Te$_4$ nanosheets with controlling thickness have been synthesized by chemical deposition approach [22]. Motivated by both theoretical predication and recent experimental realization, we set out to study the spin transport properties of Cr$_3$X$_4$ monolayers in order to explore their potential applications in spintronic devices. We shown that Cr$_3$X$_4$ monolayers are excellent candidates for applying in spin-based devices such as spin filter, spin diode and spin valve.

## II. DEVICE MODEL AND COMPUTATIONAL DETAILS

Figures 1(a) and (b) show the top and side view of monolayer Cr$_3$X$_4$ (X=Se, Te). Bulk Cr$_3$X$_4$ (X=Se, Te) is FM layered materials with space group *P-3m1*, each layer is composed of seven atomic layers with the sequence of X-Cr$_1$-X-Cr$_2$-X-Cr$_1$-X, as shown in Figure 1(b) [23,24]. To investigate the spin transport properties, we designed device based on FM Cr$_3$X$_4$ (X=Se, Te) monolayers, as shown in Figures 1(c) and (d), the top and side view of the device model, in which the transport direction belongs to the armchair configuration of the monolayers. The device can be divided into three parts: left electrode, scattering region, and right electrode. We considered two different initial magnetization configuration when calculating the transport properties, which are a parallel configuration with the same spin orientation of left and right electrodes while an antiparallel configuration with the opposite spin orientation of left and right electrodes.

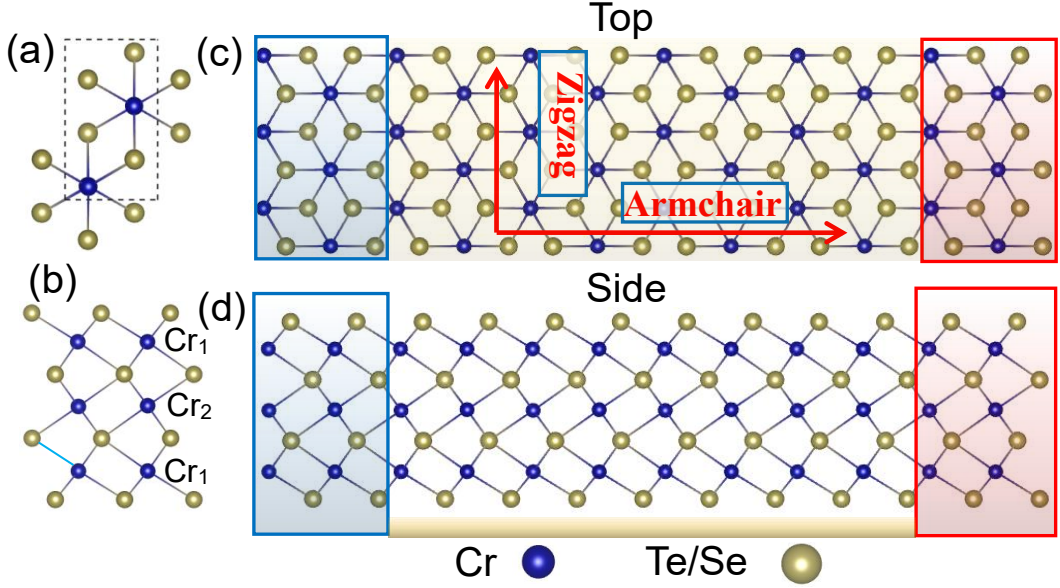

FIG. 1. (a) and (b) show the top and side view of $Cr_3X_4$ (X=Se, Te) monolayer. (c) Top view of schematic illustrations of device based on the $Cr_3X_4$(X=Te, Se) monolayer. The armchair direction is the transport direction. (d) Side view of schematic illustrations of device based on the $Cr_3X_4$(X=Te, Se). The left and right electrodes are labeled with blue and red squares, respectively.

Our calculations were performed by utilizing the density function theory (DFT) combined with non-equilibrium Green's function (NEGF) approach, as implemented in Atomistix ToolKit (ATK) package [25,26]. The spin-dependent generalized gradient approximation (SGGA) with Perdew-Burke-Ernzerhof (PBE) as the exchange-correlation function was adopted [27]. A vacuum of 20Å perpendicular to layer was placed in order to avoid the interaction between periodic layers. The cutoff energy of 150 Ry was set for all tasks. The structural optimization was converged when the force and stress tolerance are less than 0.01 eV/Å and 0.001 eV/Å$^3$, respectively. The self-consistent continues until the energy difference becomes less than $10^{-5}$ eV. The *k*-point grids used for band structure and transport calculations are 1×11×9 and 1×9×100, respectively. The current under bias voltage or temperature gradient can be obtained according to the Landauer-Buttiker formula [28]. The chemical potential of left/right electrode is $\mu_{L/R}=E_F\pm eV_b/2$ with the Fermi level at zero eV.

### III. RESULTS AND DISCUSSION

According to Zhang *et al*. [11], the FM order of $Cr_3X_4$ (X=Se, Te) monolayers primarily arises from the stronger interlayer double-exchange interaction. There are two different oxidation-state Cr atoms, which are labeled as $Cr_1$ and $Cr_2$, and shown in Figure 1(b). Based on previous calculation [11], $Cr_1$ atom loses more electron than $Cr_2$ atom, leading to different $Cr_1^{3+}$ and $C_2^{2+}$ ions. The double-exchange interaction occurs in the connected $Cr_2^{2+}$-X-$Cr_1^{3+}$ atomic chain, in which the vacant orbit of X atom can be filled by a spin-up or spin-down electron from $Cr_2^{2+}$ when it gives up the spin-up or spin-down electron to $Cr_1^{3+}$, leading to a certain spin electrons hopping between the neighboring Cr atoms. Calculated density of states (DOS) of $Cr_3X_4$ (X=Se, Te) monolayers are shown in Figure 2, one can notice that both are FM half-metals with the metallic spin up states and insulating spin down states. As one can see from Figures 2(b) and (d), the spin up states are dominantly contributed by the strongly hybridized Cr-$d$ and X-$p$ orbitals while the spin down states mainly originate from X-$p$ orbitals, which are consistent with previous report [11]. The large spin splitting and the $p$-$d$ hybridization between Cr and X atoms make $Cr_3X_4$ (X=Se, Te) monolayers half-metals.

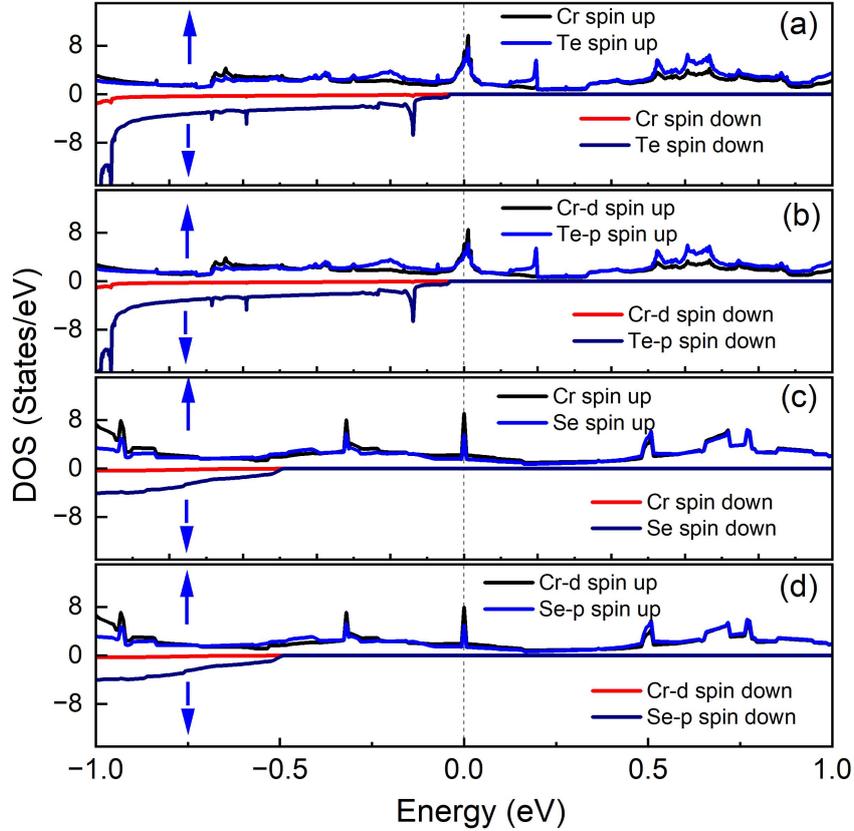

FIG. 2. (a) and (b) are projected density of states (PDOS) of $Cr_3Te_4$. (c) and (d) are PDOS of $Cr_3Se_4$. Up and down arrows represent the spin-up and spin-down polarization, respectively.

Now, let us first discuss the spin transport properties of $Cr_3Te_4$ monolayer. Figure 3(a) shows the calculated band structure of monolayer $Cr_3Te_4$ along the transport direction. In accordance with DOS, the metallic spin up channel and insulating spin down channel can be clearly distinguished. In contrast to most half-metals, it is interesting for $Cr_3Te_4$ monolayer that the spin up states occur within whole energy region (-1eV to 1eV) while spin down states only exist below and very close to the Fermi level. It is important that this unique spin polarized band structure is a robust platform for realizing a spin filter and spin diode [14,29,30]. We then calculated the spin-dependent transmission spectrum and current under a initial magnetization of parallel configuration (PC) and antiparallel configuration (APC). Calculated transition spectrum for PC and APC device under zero bias are shown in Figures 3(c) and (d), there is only spin up transport channel in PC. For APC one, although there is no transport channel under zero bias, the two opposite spin channels are highly symmetrical and very close to the Fermi level, which means that it is easy to generate spin current by applying a bias voltage.

Spin-dependent currents with respect to bias voltage from -0.5 V to 0.5 V are shown in Figures 3(g) and (h). In PC case, it is clear found that the spin down current is completely inhibited while the spin up current increases with the increase of no matter positive or negative bias voltage, indicating an excellent spin filtering effect (SFE). For APC one, the spin down current is suppressed and spin up current increases markedly within the positive bias voltage. In contrast, one can find the inhibited spin up current and rapidly increased spin down current within the negative bias voltage. In comparison to PC, the device exhibits a dual SFE under APC situation. More importantly, when taking a look at individual spin channel, it is interesting to find that the spin up current remains zero under negative bias voltage but increases rapidly within positive one, in turn, the spin down current remains zero under positive bias voltage but increases rapidly within negative one. As a result, we are able to define a dual spin diode effect (SDE) in APC situation. Moreover, the opposite spin currents within positive and negative bias voltage are highly symmetric due to the unique transmission spectrum with symmetric spin transport channels.

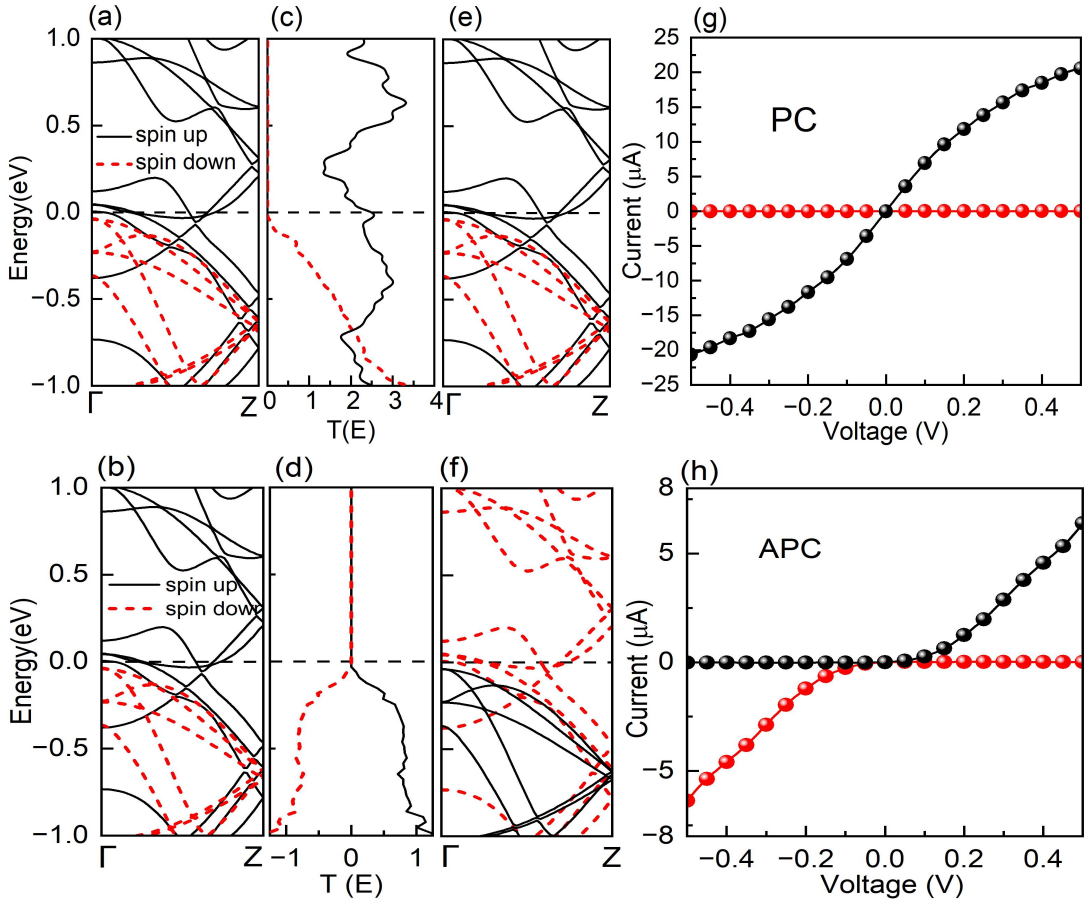

FIG. 3. The spin-dependent electronic band structure, transmission spectrum and spin-dependent I-V curve of device based on the $Cr_3Te_4$ monolayer. The above (a, c, e, g) and under (b, d, f, h) panels represent the magnetization of device in parallel configuration (PC) and antiparallel configuration (APC). (a, b) and (e, f) represent the band structure of the left and right electrodes, respectively. Black dotted lines indicate Fermi levels.

It is better to understand the SFE and SDE from the band structure and chemical potential of left and right electrodes. The Fermi level of left (right) electrodes will increase (decrease) when positive bias voltage is applied while decrease (increase) under negative bias voltage. In PC case, the band structures of left and right electrodes are the same (see Figs. 3(a) and (e)). Therefore, no matter under positive or negative bias voltage, the spin down bands of left and right electrodes can not cross the Fermi level simultaneously. Meanwhile, the spin up bands can cross the Fermi level at the same time, providing only a spin up transport channel and hence the perfect SFE of the device. In APC situation, the band structures of left and right electrodes are spin reversed (see Figs. 3(b) and (f)). When applying positive bias voltage, although the Fermi level of right electrode decreases and both spin up and spin down bands cross the Fermi level, the Fermi level of left electrode increases and provides only spin up channel, leading to only spin up current flowing through the device. Under negative bias voltage, the spin up channel is mismatched and only the spin down bands can cross the Fermi level simultaneously, generating only spin down current. Therefore, the device in APC situation shows a dual SFE and SDE.

Actually, the magnitude of current equals to the energy integral of transmission coefficient (TE) between the bias windows. As shown in Figure 4, the energy region between the two dashed lines is the bias windows $[-V_b/2, V_b/2]$ [13,31]. It can be seen that the bias window contains only spin up TE in PC case (Fig. 4(a)), which points to the perfect SFE. With the increase of bias voltage, the enlarged bias window leads to the increased integral area under the TE and thereby the growing spin up current. In APC case, the spin down TE remains zero and only spin up TE appears in the bias window when applying positive bias voltage, as shown in Figure 4(b). However, under negative bias voltage, as given in Figure 4(c), there is only spin down TE allowed in the bias window, with the spin up TE instead remains zero. As a result, pure spin up and spin down currents individually occurs under positive and negative bias voltage, respectively, suggesting the excellent dual SFE and SDF.

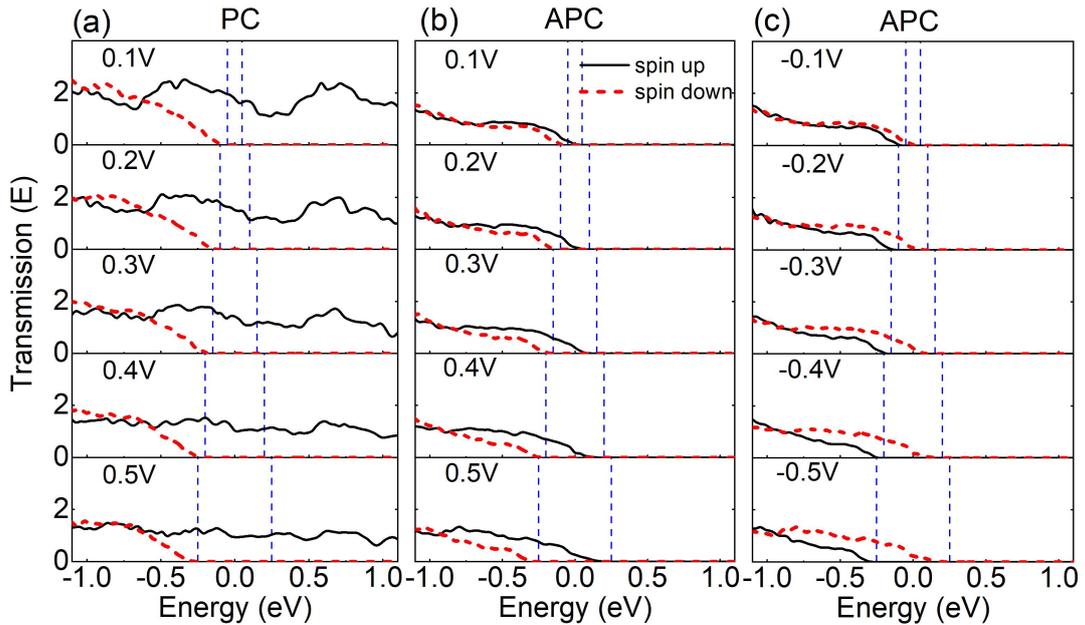

FIG. 4. The spin-dependent transmission spectrum of device based on the $Cr_3Te_4$ in the PC (a) and APC (b) within a positive bias voltage, as well as (c) in the APC within a negative bias voltage.

Now, we discuss the spin transport properties of monolayer $Cr_3Se_4$. Calculated band structure of monolayer $Cr_3Se_4$ is shown in Figure 5(a), it is crucial to find a difference compared to monolayer $Cr_3Te_4$ that the semiconducting spin down channel shifts away from the Fermi level (below -0.5 eV). The double-exchange interaction in $Cr_3Te_4$ and $Cr_3Se_4$ monolayers occurs within the electron hopping process of $Cr_2^{2+}$ ($3d$)→Te ($5p$)→$Cr_1^{3+}$ ($3d$) and $Cr_2^{2+}$ ($3d$)→Se ($4p$)→$Cr_1^{3+}$ ($3d$), respectively. In this case, the $p$-$d$ hybridization should be stronger in $Cr_3Se_4$ monolayer than in $Cr_3Te_4$ monolayer, which can be also illustrated by comparing the DOS, as given in Figure 2. The $3d$ electrons of Cr in $Cr_3X_4$ (X=Se, Te) monolayers partially occupy the splitting states with spin up orientation. As a result, the hopping electrons during the double-exchange processes belong to spin up states, leading to metallic spin up channel and semiconducting spin down channel. In addition, spin down channel mostly arises from Te $5p$ and Se $4p$ states for monolayer $Cr_3Te_4$ and $Cr_3Se_4$, respectively (see Figs. 2(b) and (d)). One can also notice that the $p$-$d$ hybridization is vanishing in spin down channel due to the difficult hopping process of spin down electrons in the double-exchange interaction. Accordingly, the lower energy of $4p$ states leads to the spin down channel shifting down from the Fermi level in monolayer $Cr_3Se_4$.

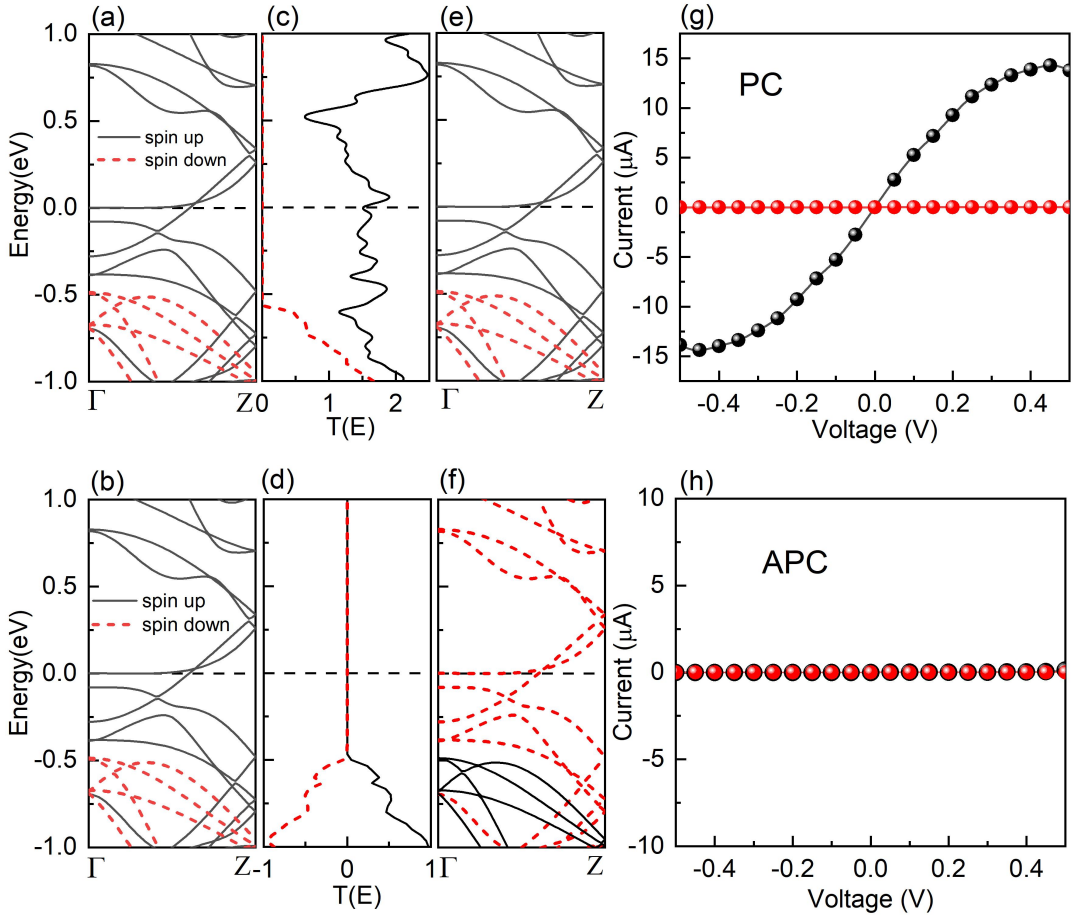

FIG. 5. The spin-dependent electronic band structure, transmission spectrum and spin-dependent I-V curve of device based on the $Cr_3Se_4$ monolayer, the above and under panels denote the magnetization in the PC (a, c, e, g) and APC (b, d, f, h), respectively. (a, b) and (e, f) represent the band structure of the left and right electrodes, respectively. Black dotted lines indicate Fermi

levels.

Figures 5(c) and (d) show the calculated TE of monolayer $Cr_3Se_4$ based device in PC and APC, respectively. For PC case, the band structure of left and right electrodes are the same (see Figs. 5(a) and (e)), only spin up channel occurs in the TE (see Fig. 5(c)). No matter positive or negative bias voltage is applied, the spin up bands of both electrodes will cross the Fermi level, bringing about the spin up current in the device, as shown in Figure 5(g). However, there is no spin down current because the spin down bands of both electrodes can not cross the Fermi level under limited bias voltage. Consequently, the device in PC also remains an excellent SFE. For APC case, the band structure of left and right electrodes are spin reversed. In contrast to monolayer $Cr_3Te_4$, the two symmetric spin transport channels are far from the Fermi level (below -0.5 eV) due to the shifting of spin down bands in monolayer $Cr_3Se_4$, as plotted in Figure 5(d), which means that it is difficult to activate the transport channels within limited bias voltage. In Figure 5(h), it is clear to see that there is no current for any spin channels within positive or negative bias voltage, as both the spin up and spin down bands of left and right electrodes can not cross the Fermi level simultaneously. Consequently, the current in monolayer $Cr_3Se_4$ based device arises in PC while disappears in APC, indicating an excellent spin valve effect. Figure 6(a) shows the calculated magnetoresistance (MR) ratio, which is defined as MR=$(I_P-I_{AP})/I_{AP}$, $I_P$ and $I_{AP}$ are the total current of PC and APC, respectively. The maximal MR ratio reaches $2\times10^3$, which is comparable to that of CoFeMnSi/GaAs/CoFeMnSi MTJ [13].

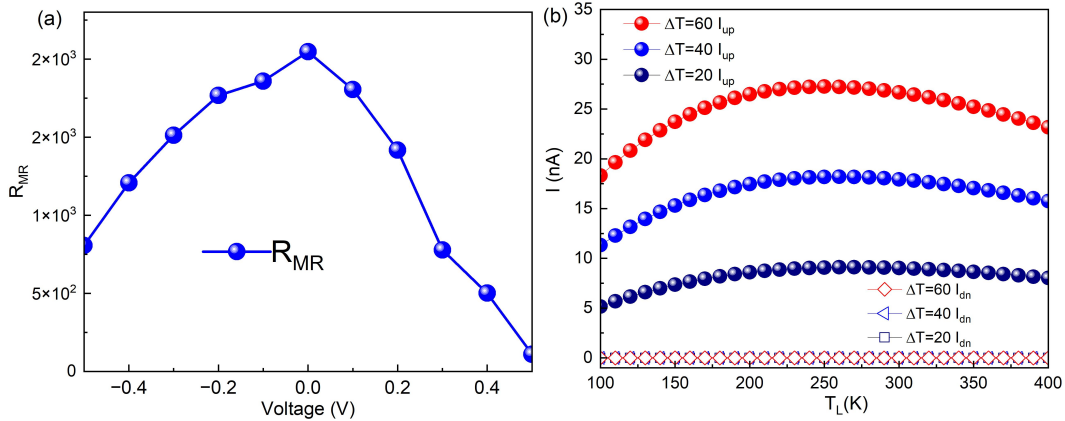

FIG. 6. (a) The magnetoresistance ($R_{MR}$) of $Cr_3Se_4$ homojunction versus the bias voltage. (b) Calculated spin-dependent currents of $Cr_3Se_4$ junction under a temperature gradient as a function $T_L$ at different $\triangle T$.

It is also interesting to find a negative differential resistance effect (NDRE) of monolayer $Cr_3Se_4$ based device in PC case. As shown in Figure 5(g), the spin up current increases with the increase of bias voltage and reaches a peak value around $\pm0.45$ V, then decreases with the further increase of voltage, indicating a significant NDRE. The NDRE can be understood from the TE of spin up channel above the Fermi level (see Fig. 5(c)), where the TE has a peak near the Fermi level and then declines with the increase of energy. Therefore, with the increase of bias voltage, the energy integral of TE between the bias window increases first and decreases later (around $\pm 0.45$ V), leading to the NDRE in the junction. In addition, we also considered the transport properties of monolayer $Cr_3Se_4$ in PC by applying a temperature gradient instead of bias voltage.

The left and right electrodes are defined as low and high temperature end, respectively. When the temperature gradient is applied, both the electrons and holes will be excited and move from right to left electrode, generating electron and hole current flowing in opposite direction [32-34]. Figure 6(b) shows the calculated thermally induced spin current as a function of temperature of left electrode. A thermally induced SFE is found since there is only spin up channel in TE, and the positive spin up current indicates the dominated contribution of spin up electrons as arising from the peak TE above the Fermi level. As a result, the spin up current flows from left electrode to right electrode. Furthermore, a thermally induced NDRE is also observed due to decline of TE above the Fermi level.

## IV. CONCLUSION

In summary, motivated by recently experimental growth of ferromagnetic $Cr_3Te_4$ nanosheet, we designed devices based on $Cr_3X_4$ (X=Se, Te) monolayers to explore their potential application in spintronics. On the basis of density function theory combined with non-equilibrium Green's function method, we found that monolayer $Cr_3Te_4$ is a robust platform to realize a spin filter and dual spin diode. In contrast, due to the shifting of spin down transport channel, monolayer $Cr_3Se_4$ can be applied as a spin valve with a high magnetoresistance ratio (up to $2 \times 10^3$). A negative differential resistance effect was discovered in monolayer $Cr_3Se_4$ based device for PC case, no matter under bias voltage or temperature gradient. All results were explained from the calculated spin-dependent band structure and transmission spectrum. Our theoretical results highlight $Cr_3X_4$ (X=Se, Te) monolayers as the promising candidates to realize robust spin filter, spin diode and spin valve.

**ACKNOWLEDGMENTS**

This work is supported by the National Natural Science Foundation of China with Grants No.11804040.

[1] T. Thuy. Hoang, S. H. Rhim, and S. C. Hong, Phys. Rev. Mater. **6**, 055001 (2022).

[2] R. Meng, M. Houssa, K. Iordanidou, G. Pourtois, V. Afanasiev, and A. Stesmans, Phys. Rev. Mate. **4**, 074001 (2020).

[3] D. Wang, M. Shaikh, S. Ghosh, and B. Sanyal, Phys. Rev. Mater. **5**, 054405 (2021).

[4] T. Song, X. Cai, M. W. Y. Tu, X. Zhang, B. Huang, N. P. Wilson, K. Seyler, L. Zhu, T. Taniguchl, K. Watanabe, M. A. Mcguire, D. H. Cobden, D. Xiao, W. Yao, and X. Xu, Science. **360**, 1214-1218 (2018).

[5] K. Szulc, P. Graczyk, M. Mruczkiewicz, G. Gubbiotti, and M. Krawczyk, Phys. Rev. Appl. **14**, 034063 (2020).

[6] F. Casanova, A. Sharoni, M. Erekhinsky, and I. K. Schuller, Phys. Rev. B. **79**, 184415 (2009).

[7] Q. Liu, Y. Guo, and A. J. Freeman, Nano. Lett.**13**, 5264-5270 (2013).

[8] M. Matsuo, Y. Ohnuma, T. Kato, and S. Maekawa, Phys. Rev. Lett. **120**, 037201 (2018).

[9] F. Wu, C. Huang, H. Wu, C. Lee, K. Deng, E. Kan, and P. Jena, Nano. Lett. **15**, 8277-8281 (2015).

[10] Y. Feng, X. Wu, J. Han, and G. Gao, J. Mater. Chem. C. **6**, 4087-4094 (2018).

[11] X. Zhang, B. Wang, Y. Guo, Y. Zhang, Y. Chen, and J. Wang, Nanoscale. Horizon. **4**, 859-866

(2019).

[12] Z. Chen, X. Fan, Z. Shen, Z. Luo, D. Yang, and S. Ma, J. Mater. Sci. **55**, 7680-7690 (2020).

[13] J. Han, Y. Feng, K. Yao, and G. Y. Gao, Appl. Phys. Lett. **111**, 132402 (2017).

[14] J. Li, G. Gao, Y. Min, and K. Yao, Phys. Chem. Chem. Phys. **18**, 28018-28023 (2016).

[15] D. D. Wu, H. H. Fu, Q. B. Liu, G. F. Du, and R. Wu, Phys. Rev. B. **98**, 115422 (2018).

[16] Y. Feng, X. Wu, L. Hu, and G. Gao, J. Mater. Chem. C. **8**, 14353-14359 (2020).

[17] T. Song, M. W. Y. Tu, C. Carnahan, X. Cai, T. Taniguchi, K. Watanabe, M. A. Mcguire, D. H. Cobden, D. Xiao, W. Yao, and X. Xu, Nano. Lett. **19**, 915-920 (2019).

[18] G. Cheng, L. Lin, L. Zhenglu, J. Huiwen, S. Alex, X. Yang, C. Ting, B. Wei, W. Chenzhe, W. Yuan, Z. Q. Qiu, R. J. Cava, G. L. Steven, X. Jing and Z. Xiang, Nature. **546**, 265-269 (2017).

[19] B. Huang, G. Clark, E. Navarro-Moratalla, D. R. Klein, R. Cheng, K. L. Seyler, D. Zhong, E. Schmidgall, M. A. McGuire, D. H. Cobden, W. Yao, D. Xiao, P. Jarillo-Herrero, and X. Xu, Nature. **546**, 270-273(2017).

[20] Z. Fei, B. Huang, P. Malinowski, W. Wang, T. Song, J. Sanchez, W. Yao, D. Xiao, X. Zhu, A. May, W. Wu, D. H. Cobden, J. Chu and X. Xu, Nat. Mater. **17**, 778-782 (2018).

[21] X. Zhou, B. Brzostowski, A. Durajski, M. Liu, J. Xiang, T. Jiang, Z. Wang, S. Chen, P. Li, Z. Zhong, A. Drzewiński*, M. Jarosik, R. Szczęśniak, T. Lai, D. Guo, and D. Zhong, J. Phys. Chem. C. **124**, 9416-9423(2020).

[22] B. Li, X. Deng, W. Shu, X. Cheng, Q. Qian, Z. Wan, B. Zhao, X. Shen, R. Wu, S. Shi, H. Zhang, Z. Zhang, X. Yang, J. Zhang, M. Zhong, Q. Xia, J. Li, Y. Liu, L. Liao, Y. Ye, L. Dai, Y. Peng, B. Li, and X. Duan, Mater Today. doi.org/10.1016/j.mattod.2022.04.011.

[23] J. Dijkstra, H. H. Weitering, C. F. Van Bruggen, C. Haas, and R. A. De Groot, J. Phys-Condens. Mat. **1**, 9141(1989).

[24] M. Yamaguchi, and T. Hashimoto, J. Phys. Soc. Jpn. **32**, 635-638(1972).

[25] J. Taylor, H. Guo, and J. Wang, Phys. Rev. B. **63**, 245407 (2001).

[26] M. Brandbyge, J. L. Mozos, P. Ordejón, J. Taylor, and K. Stokbro, Phys. Rev. B. **65**, 165401 (2002).

[27] J. P. Perdew, K. Burke, and M. Ernzerhof, Phys. Rev. Lett. **77**, 3865 (1996).

[28] Y. Imry, and R. Landauer, Rev. Mod. Phys. **71**, S306 (1999).

[29] L. Hu, X. Wu, Y. Feng, Y. Liu, Z. Xu, and G. Gao, Nanoscale. **14**, 7891-7897(2022).

[30] M. Zeng, L. Shen, M. Zhou, C. Zhang, and Y. Feng, Phys. Rev. B. **83**, 115427(2011).

[31] Y. Feng, N. Liu, and G. Gao, Appl. Phys. Lett. **118**, 112407(2021).

[32] X. Tan, L. Ding, G. F. Du, and H. H. Fu, Phys. Rev. B. **103**, 115415(2021).

[33] H. H. Fu, G. F. Du, D. D. Wu, Q. B. Liu, and R. Wu, Phys. Rev. B. **100**, 085407(2019).

[34] D. D. Wu, Y. T. Ji, G. F. Du, X. Y. Yue, Y. Y. Wang, Q. J. Li, X. F. Sun, and H. H. Fu, J. Mater. Chem. C. **10**, 3188-3195(2022).